\documentclass[prb,superscriptaddress,twocolumn,floatfix]{revtex4-1}
\usepackage{graphicx,color}
\usepackage{amsmath,amssymb,bm}
\usepackage{enumerate}
\usepackage{braket}		

\usepackage{amsmath}

\begin{document}
\title{Density-Matrix Renormalization Group study of \\ Many-body Localization in Floquet Eigenstates}

\author{Carolyn Zhang}
\affiliation{Max Planck Institute for the Physics of Complex Systems, Dresden 01187, Germany}
\affiliation{Yale University, New Haven, CT 06520, USA}
\author{Frank Pollmann}
\affiliation{Max Planck Institute for the Physics of Complex Systems, Dresden 01187, Germany}
\author{S. L. Sondhi}
\affiliation{Max Planck Institute for the Physics of Complex Systems, Dresden 01187, Germany}
\affiliation{Physics Department, Princeton University, Princeton, NJ 08544, USA}
\author{Roderich Moessner}
\affiliation{Max Planck Institute for the Physics of Complex Systems, Dresden 01187, Germany}

\begin{abstract}
We generalize the recently introduced Density-Matrix Renormalization Group (DMRG-X) [Khemani et al, PRL 2016] algorithm to obtain Floquet eigenstates of one-dimensional, periodically driven many-body localized systems. 
This generalization is made possible by the fact that the time-evolution operator for a period can be efficiently represented using a matrix-product operator. 
We first benchmark the method by comparing to exact diagonalization for small systems. 
We then obtain Floquet eigenstates for larger systems and show unambiguously that the characteristic area-law scaling remains robust.
\end{abstract}
\maketitle

\section{Introduction}

Quantum many-body systems at sufficiently strong disorder can exhibit a so-called \textit{many-body localized} (MBL) phase as shown in by Basko et al. in Ref.~\onlinecite{Basko:2006hh} (see also Refs.~\onlinecite{fleishman1980interactions,Gornyi:2005})
This dynamical phenomenon is an extension of Anderson localization\cite{Anderson:1958vr} to interacting systems and involves a breakdown of statistical mechanics wherein MBL quantum systems do not thermalize.\cite{Nandkishore2014} 
MBL systems can also be characterized by the properties of their highly excited eigenstates and it was argued \cite{Pal:2010gr} that such eigenstates would have only local entanglement and thus obey the area law\cite{Eisert:2010hq}.
This conjecture has further been corroborated  by subsequent numerical studies examining the behavior of entanglement entropy in highly excited eigenstates.\cite{Bauer:2013jw,Luitz:2015,Kjaell:2015df}
In contrast to the area law in eigenstates, a logarithmic growth of entanglement without bounds as a function of time is found following a global quench.\cite{Znidaric:2008cr,Bardarson:2012gc,Serbyn:2013he,Vosk:2013kt} Interestingly, even in the MBL regime it is possible to have a sharp definition of phases and phase transitions based on the notion of eigenstate order \cite{Huse:2013bw,Pekker:2013vt,Vosk2014,Chandran:2013uz,Bahri:2013ux}.
A description of MBL systems was introduced based on the existence of a complete set of commuting local integrals of motion (also called ``l-bits'') which are believed to exist in systems in which \emph{all} many-body eigenstates are localized \cite{Serbyn:2013cl,Huse:2013uc,Huse:2014uy,Chandran:2015df,Ros2015,Huse:2014uy} and this captures all of the features mentioned above. 

The question of how quantum many-body systems behave when exposed to periodic driving with a  Hamiltonian $H(t)=H(t+T)$ is currently a focus in the field of non-equilibrium physics.\cite{Lazarides:2014ie,Lazarides:2015jd,abanin2015effective,DAlessio:2014fg,Ponte:2015dc,Ponte:2015hm,Lazarides:2014ie,DAlessio:2014fg,Ponte:2015hm,Eckardt:2008fg,Genske:2015iq,Ponte:2015dc,Lazarides:2015jd,rehn2016periodic}
With sufficient disorder, it is expected that periodically driven (or Floquet) systems can exhibit a dynamical quantum phase transition from a localized to an extended, ``infinite-temperature'' phase as function of the driving.
The properties of a Floquet system are determined by the eigenstates $|\psi_{n}\rangle$ of its time evolution operator $U_T$ for one period. 
In its eigenbasis, the unitary can be expressed as $U_T=e^{i\theta_n}|\psi_n\rangle\langle\psi_n|$ with $\theta_n\in[-\pi,\pi)$.
We can define a so called Floquet Hamiltonian $H_{F}$ with eigenvalues $\theta_n$ and corresponding eigenstates $|\psi_n\rangle$ such that $U_T = e^{iTH_F}$.
Like the MBL phase in static systems, the Floquet MBL phase is characterized by the locality of the eigenstates, including an area law and the conjectured existence of quasi local integrals of motion.\cite{Abanin20161}
Finally, Floquet MBL systems can also exhibit phases in which the Floquet eigenstates exhibit different forms of order \cite{Khemani16},
some unique to the driven setting. A spate of recent work has identified a large set of such phases with symmetry breaking and 
topological order.\cite{Keyserlingk16,Keyserlingk16b,Roy2016,Else2016,Potter2016,Keyserlingk16c}

To understand the phenomenon of MBL in both in static and driven systems, we must solve a quantum many-body problem. 
Due to the exponentially growing Hilbert space, it is prohibitively expensive to diagonalize systems larger than $\sim$$20$ spins. 
In Ref.~\onlinecite{khemani2016obtaining}, a generalization of the density matrix renormalization group  method \cite{White:1992,Schollwock2011} (DMRG-X) was introduced that  obtains \emph{highly excited} eigenstates of MBL systems to machine precision at moderate to large disorder in a time that scales only polynomially with $N$ (see also related works in Refs.~\onlinecite{Pekker:2014ux,pollmann16,lim15,yu2015finding,serbyn16}).
The underlying principle is that quantum states that are only slightly entangled (i.e., area law states) can be efficiently represented using matrix-product states (MPS)\cite{Fannes-1992,Schuch:2008ty,PhysRevLett.114.170505}.  Furthermore, the DMRG-X method explicitly takes advantage of the local spatial structure of the  eigenstates.

In this paper, we generalize the DMRG-X method to find the eigenstates of Floquet MBL systems. 
An important insight for this generalization is that the Floquet operator $U_T$ can be compressed into a matrix-product operator (MPO) representation.\cite{verstraete}
This compression is performed iteratively using a variant of the time-evolving block decimation (TEBD) algorithm.\cite{vidal2004}
We  first benchmark the method by comparing with exact diagonalization (ED) results for small system sizes. 
By  going to systems far beyond ED, we can then unambiguously show that the MBL phase remains robust over a range of parameters.  
In particular, we measure the entanglement entropy of the eigenstates and show that the system studied exhibits the characteristic area law. 

The remainder of this paper is organized as follows. In Sec.~\ref{sec:algo} we introduce and discuss the TEBD + DMRG-X algorithm to find highly excited states of Floquet MBL systems. 
In Sec.~\ref{sec:binary}, we introduce a binary drive model and briefly review a previous ED study. 
We benchmark our algorithm with exact results for small systems in Sec.~\ref{sec:ed} and then show results for large systems in Sec.~\ref{sec:big}.
We finally conclude with a summary and outlook in Sec.~\ref{sec:disco}.

\section{Algorithm}\label{sec:algo}
The algorithm to efficiently find MPS representations of Floquet eigenstates introduced in this paper is a generalization of the DMRG-X algorithm for static MBL systems. 
A key difference is that we first need to construct an MPO representation of the the Floquet operator.  
\subsection{Efficient representation of states and operators}
A general quantum state $|\Psi\rangle$ for a one-dimensional system of $N$ sites can be written in a canonical MPS form \cite{VidalCanonical} such that
\begin{eqnarray}
	\ket{\Psi} = \sum_{\{j_n\}}
		\Gamma^{[1]j_1} \Lambda^{[1]} \Gamma^{[2]j_2} \Lambda^{[2]} \cdots \Lambda^{[N-1]} \Gamma^{[L]j_N}| j_1, \ldots ,j_{N} \rangle.\nonumber\\
 \label{eq:mps}
\end{eqnarray}
Here,  $\Gamma^{[n]j_n}$  are $\chi_{n} \times \chi_{n+1}$ matrices and $|j_n\rangle$ with $j_n=1,\dots,d$ is a basis of local states at site $n$ (for a spin 1/2 system, $d = 2$).  
The matrices at the boundaries,  $n=1$ and $n=N$, are vectors so that $\chi_1 = \chi_{N+1} = 1$ and the product over all matrices gives a complex number, which is the amplitude of $|\Psi \rangle$ on the basis state $|j_1 \cdots j_N \rangle$. 
The  $\Lambda^{[n]}$ are diagonal matrices in which the entries correspond to the Schmidt values for a decomposition at bond $n$ into a sum of $\chi$ orthogonal wave functions $\ket{\alpha}_{[1,\dots, n]}$ and $\ket{\alpha}_{[n+1,\dots, N]}$ to the left/right of the bond. 
The state then takes the form
\begin{eqnarray}
	\ket{\Psi} &= \sum_{\alpha=1}^\chi \Lambda^{[n]}_\alpha \ket{\alpha}_{[1,\dots,n]} \otimes \ket{\alpha}_{[n+1,\dots,N]}.
\end{eqnarray}
The wave functions $|\alpha\rangle_{L/R}$ are formed by simply multiplying all matrices to the left and right, respectively. The index $\alpha$ is the dangling index on the left / right of the matrices at site $n$ / $n+1$.

\begin{figure}
	\begin{center}
	\includegraphics[width=8cm]{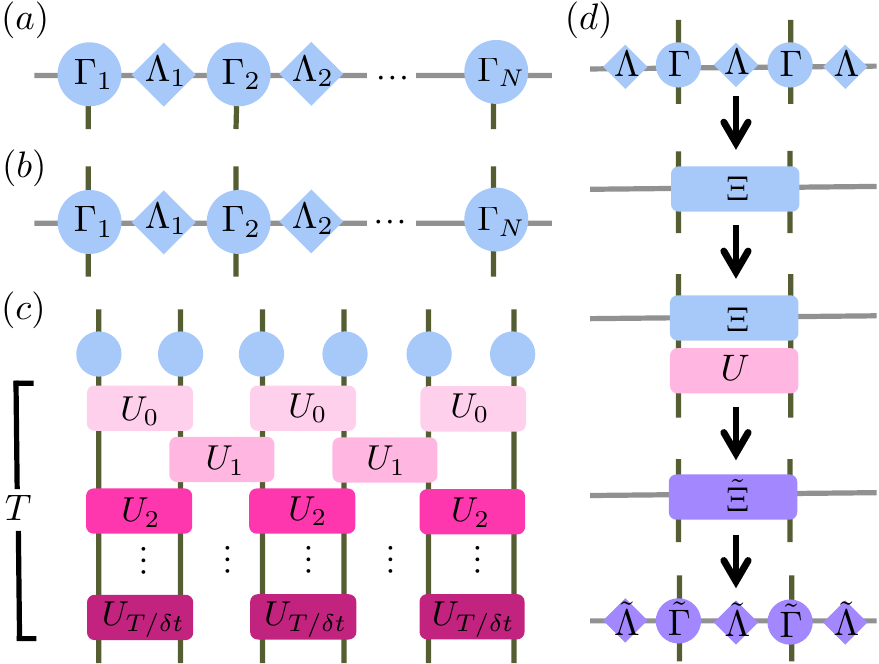}
	\caption{Schematic tensor representation: (a) and (b) exemplify a generic MPS and MPO respectively, both in canonical form where $\Lambda_n$ are positive, real, square diagonal matrices containing the Schmidt values.  (c) overviews the TEBD method for creating the Floquet MPO.  Successive two-site unitary operators are applied, from $U_0=U(0,\delta t)$ to $U_{T/\delta t}=U(T-\delta t,T)$.  (d) focuses on the evolution of one site over time $\Delta$, showing contraction of two sites, application of a unitary operator, and singular value decomposition (SVD) . \label{fig:mpsmpo}}
	\end{center}
\end{figure}

A  similar idea applies to the space of linear operators which can in turn be represented in terms of MPOs taking the form
\begin{eqnarray}
	O= \sum_{\{j_n\}\{j_n'\}}
		\Gamma^{[1]j_1j_1'} \Lambda^{[1]} \Gamma^{[2]j_2j_2'} \Lambda^{[2]} \cdots \Lambda^{[N-1]} \Gamma^{[N]j_N}\times\nonumber\\
		\times | j_1, \ldots ,j_{N} \rangle\langle j_1', \ldots ,j_{N}' |.\nonumber\\
 \label{eq:mps}
\end{eqnarray}
Here the only difference compared to MPS is that we added one physical index to the tensors.
Further, the norm $\mathcal{N} = \sum_{\alpha}\Lambda_{\alpha}^2$ corresponds to the two-norm of the linear operator.
In the following, we use a graphical representation of the MPSs and MPOs as shown in Fig.~\ref{fig:mpsmpo}(a) and (b). 

\subsection{Floquet MPO}
We now discuss how to compress a Floquet operator $U_T$ into an MPO form assuming that the terms in the driven Hamiltonian are short-ranged. 
%

The Floquet time evolution operator for one period $T$ of driving is given by
\begin{equation}
U_T = \mathcal{T}\exp\left\{-i\int_0^TH(t)dt\right\},
\end{equation}
where $\mathcal{T}$ denote the time ordering operator. The compression of $U_T$  is performed using the time-evolving block decimation algorithm.\cite{vidal2004} While originally proposed to evolve MPS, it can analogously be used to compress a unitary into an MPO.
The algorithm uses Trotter-Suzuki decomposition, which approximates the exponent of a sum of operators with a product of exponents of the operators. 
For example, the first order expansion reads
\begin{equation}
	e^{(V+W)\delta} = e^{V\delta}e^{W\delta} + \mathcal{O}(\delta^2).
	\label{eq:ST1}
\end{equation}
Let us now assume that the Hamiltonian is a sum of two-site operators of the form $H(t)=\sum_n h^{[n,n+1]}(t)$ and decompose it as a sum 
\begin{align}
	H &= H_{\rm odd}(t) + H_{\rm even}(t) \notag\\
	&= \sum_{n\; {\rm odd}} h^{[n,n+1]}(t) + \sum_{n\; {\rm even}} h^{[n,n+1]}(t). 
\end{align}
Note that each term $H_{\rm odd}$(t) and $H_{\rm even}$(t) consists of a sum of commuting operators.
We now divide the time into small time slices $\delta t$  and consider a time evolution operator $U(t,t+\delta t)$. 
Using, as an example, the first order decomposition~\eqref{eq:ST1}, the operator $U(t,t+\delta t)$ can be expanded into products of two-site unitary operators
\begin{equation}
	\left[\prod_{n\; {\rm odd}} U^{[n,n+1]}(t,t+\delta t)  \right]\left[\prod_{n\; {\rm even}} U^{[n,n+1]}(t,t+\delta t)  \right],
	\label{TimeEvol}
\end{equation}
where
\begin{eqnarray}
	U^{[n,n+1]}(t,t+\delta t)=e^{-i \, \delta t \, h^{[n,n+1]}(t)}
\end{eqnarray}
The successive application of these two-site unitary operators for one period $T$ is the main part of the algorithm. 
A pictorial representation of the algorithm is given in Fig.~\ref{fig:mpsmpo}(c) and (d). 
This simple decomposition corresponds to a first order Trotter decomposition with  total error $\mathcal{O}(\delta t)$. 
It is straight forward to perform higher order decompositions.\cite{Schollwock2011}
To start, a simple product operator is formed by a tensor product of identity operators. 
Then the two-site gates $U^{[i,i+1]}(t,t+\delta t)$ are successively applied alternating even and odd bonds. 
Each application of a two-site gate follows the same procedure:
\begin{enumerate}[(i)]
\item Contract two neighboring tensors $\Gamma^{[n]}$ and  $\Gamma^{[n+1]}$  with adjacent $\Lambda$'s to form a mixed representation of $\Xi^{ii';jj'_{\alpha\beta}}$ in terms of the local operators $|ij\rangle\langle i'j'|$ and Schmidt states to the left and right. 
\item Apply the two site gate $U^{[n,n+1]}(t,t+\delta t)$ by contracting over two physical indices.
\item Group the legs left and right of the center and perform a singular value decomposition (SVD) of the matrix $\Xi_{ii'\alpha;jj'\beta}$ to obtain the new tensors.
\item Truncate by neglecting small singular values and renormalize the operator such that $\sum_{\alpha} (\tilde{\Lambda}_{\alpha}^{[n]})^2=\mathcal{N}.$ 
\end{enumerate}

Note that the compression of the Floquet operator is  efficient as long as the period is sufficiently short independent of the system size $N$. 
In particular, for a generic local Hamiltonian, the size of the dimension of the matrices needed to express the operator grows exponentially with time due to to the fact that degrees of freedom within a light cone get entangled.\cite{lieb:1972,kliesch2014lieb}
In special cases the growth of the bond dimension can be much slower. 
For example due to the slow growth of entanglement in MBL systems, the bond dimension required to express the unitary is expected to only grow polynomially with time. 
Further, for Hamiltonians consisting of commuting operators, the dimension is independent of the length of the driving period.

 \subsection{DMRG-X}

Once we have obtained the MPO representation of the Floquet operator $U_T$ we can use DMRG-X to variationally find its eigenstates in MPS form.
The procedure is very similar to the original DMRG-X procedure, with the difference that the Floquet operator is usually not Hermitian.
Let us now briefly review the basic concepts of the DMRG-X algorithm (details of the algorithm can be found in Ref.~\onlinecite{khemani2016obtaining}).
The DMRG-X algorithm diagonalizes the effective Floquet operator $\mathcal{U}_T$  in a mixed representation which expresses local states in the spin basis $|j_{n}\rangle\otimes|j_{n+1}\rangle$ and the environment in terms of orthogonal Schmidt states $\ket{\alpha}_{[1,\dots,n-1]} \otimes \ket{\beta}_{[n+2,\dots,N]}$.
This effective Floquet operator is generated at each step from the MPO representation of $U_T$.
The algorithm is initialized by choosing a product state that has a finite overlap with some l-bit state, e.g., for the binary drive model discussed below we choose random states in the $\sigma^z$  basis.  We then start our DMRG-X algorithm with the following local two-site update at sites [n,n+1]:
\begin{enumerate}[(i)]
 \item Construct the effective Floquet operator $\mathcal{U}_T$ in the mixed basis for sites $[n,n+1]$.
 \item Diagonalize  $\mathcal{U}_T$ and pick the eigenstate  that has \emph{maximum overlap} with the current MPS.
 \item Update the tensors on sites [n,n+1] and move to the next bond.
 \end{enumerate}

The sweeping procedure is repeated until the MPS converged to a Floquet eigenstate with the ``l-bit'' quantum number selected by the initial configuration.
To accelerate the algorithm for larger bond dimensions, we switch to a sparse matrix method once the bond dimension has reached a certain threshold.
We then find a number of $n_{x}$ eigenstates near the quasi energy $E = \log \langle\psi| \mathcal{U}_T | \psi\rangle$ using an iterative shift-invert method, with $|\psi\rangle$ being the current MPS.
We the successively increase $n_{x}$  in case none of the excited states has a significant overlap with $|\psi\rangle$.

From the eigenstates, we can then investigate signatures of MBL such as area-law entanglement entropy, fluctuations in entropy and other observables, participation ratio, and correlation functions (i.e. spin glass order parameter, which we do not explore here).  
Due to the periodic nature of Floquet quasi-energy spectrum, we assume that the properties of all the eigenstates are statistically the same and confirm this by studying eigenstates with different quasienergies.
Therefore, the behavior of a single eigenstate of $U_T$ is expected to describe all the eigenstates, and fully characterize the thermalization properties of the system.

\section{Binary drive model}\label{sec:binary}
To test the algorithm, we consider a binary drive  \cite{Ponte:2015dc,d2013many} in which the Hamiltonian dynamically switches using the following protocol. 
For a time $T_0$ a localized Hamiltonian 
\[
H_0 = \sum_{i}h_i\sigma_i^z+J_z\sigma_i^z\sigma_{i+1}^z,
\]
is applied, where the $h_i$ are uniformly distributed in $[-W,W]$. Then the delocalizing part
\[
H_1 = J_{\perp}\sum_{i}\sigma_i^x\sigma_{i+1}^x+\sigma_i^y\sigma_{i+1}^y 
\]
acts for a time $T_1$ so that the total period is $T=T_0+T_1$. 
The corresponding Floquet operator is then $U_T=e^{-iH_0T_0}e^{-iH_1T_1}$. 
Ponte et al.  studied in Ref.~\onlinecite{Ponte:2015dc} this model varying $T_1$ for fixed  $J_{\perp}=J_z=1/4$, $W=2.5$, and $T_0=1$; we adapt the same parameters for our work.
Using ED of small ($N=10-14$) system sizes, as well as evolution of local observables and entanglement (up to $N=30$), it was argued that the binary drive model exhibits an MBL transition as function of $T_1$. 
For small values of $T_1$, the system is found to be MBL while for large $T_1$ the Floquet operator is delocalized.  
Different measures of MBL resulted in ambiguous values for a critical $T_1^C$ for a phase transition, giving a $T_1^C$ between ~0.6 to 1.1.  
Furthermore, significant drifts were observed for the values obtained by ED due to finite size effects. 

\section{Comparison to exact diagonalization}\label{sec:ed}

\begin{figure}
	\includegraphics[width=8cm]{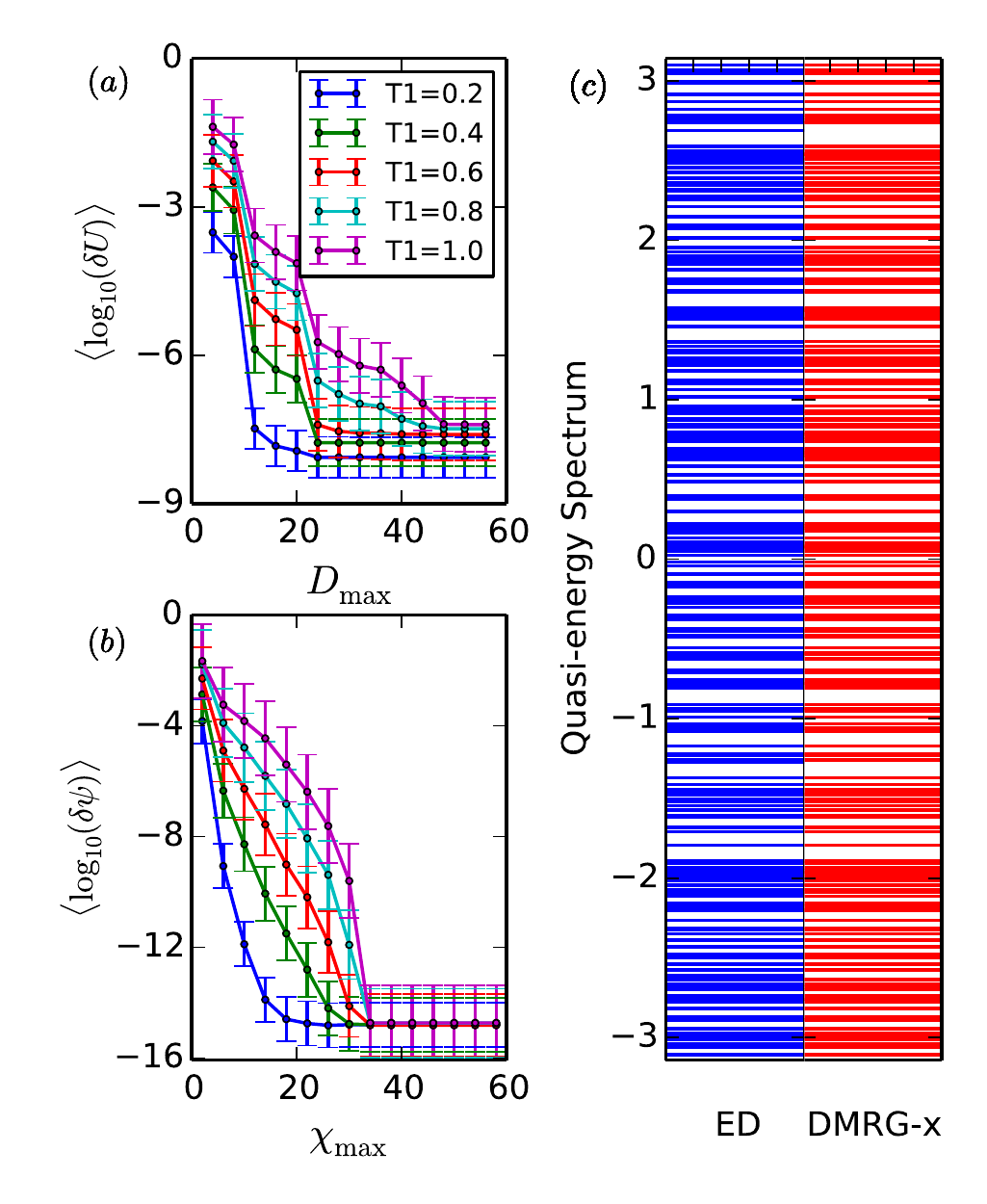}
	\caption{ Comparison with ED for a chain of length $N=10$ a timestep of $\delta t=10^{-3}$. Shown is the average over 100 different disorder realizations (a) $\delta U$ from varying $D_{\mathrm{max}}$ \, with $\chi_{\mathrm{max}}=32$ to guarantee zero error from DMRG-X.  (b) $\delta\psi$ from varying $\chi_{\mathrm{max}}$ with $D_{\mathrm{max}}$ adjusted to maintain a constant, low $\delta U$.  (c) Quasi-energy spectrum from ED vs. TEBD + DMRG-X.\label{fig:EDcompare}}	\end{figure}

We now benchmark our method by comparing with ED results for small system sizes.
The simulation contains errors due to approximations made both during the MPO compression of the unitary and the MPS representation of the eigenstates.

First we compare in Fig.~\ref{fig:EDcompare}(a) the approximated Floquet operator with the exact one for $N=10$ by plotting the norm difference 
\begin{equation}
\delta U =  \frac{\lVert U_{\mathrm{ED}}-U_{\mathrm{MPO}}\lVert_2}{\lVert U_{\mathrm{ED}}\lVert_2}
\end{equation}
for a range of $D_{\mathrm{max}}$ with $\lVert\cdot\lVert_2$ being the Frobenius norm.  $U_{\mathrm{ED}}$ is the exact unitary with dimensions $D=d^{2N}$, and $U_{\mathrm{MPO}}$ is the unitary created from contracting the truncated MPO along all the bond indices.
$U_{\mathrm{MPO}}$ is affected by Trotter error due to the finite time step in second order Trotter decomposition as well as truncation error due to truncation to a bond dimension $D_{\mathrm{max}}$.
Evidently, for  small $T_1\le 1$, a small $D_{\mathrm{max}}$$\sim$$20$ is sufficient to faithfully represent the unitary $U_{T}$ and error comes solely from Trotter decomposition.
Further, we tested that for a given small $T_1$, the required $D_{\mathrm{max}}$ does not depend on the system size (not shown).
However, as the delocalization time $T_1$ is increased, $D_{\mathrm{max}}$ must also be increased to capture the same accuracy.
A few comments are in order:
(i) The unitary $U_T$ can be efficiently represented with a rather small bond dimension even though the system is already in the extended phase as the total time $T_0 + T_1$ is short. 
(ii) For the specific binary drive used, the required $D_{\mathrm{max}}$ depends only weakly on the specific disorder realization as the disordered part $H_0$ contains only commuting terms.
(iii) While the Trotter error can easily be further reduced, we found that an error of $10^{-6}$ is sufficient to find the eigenstates up to machine precision. 
(iv) In principle it is possible to optimize the choice of the Floquet operator over the different choices of initial time to
obtain the smallest average bond dimension over the period.

Next we compare the quality of the eigenstates obtained using DMRG-X from a fully converged Floquet operator ($D_{\mathrm{max}}=40$). In this case, the only error is due to the finite bond dimension $\chi_{\mathrm{max}}$ of the MPS. 
Figure~\ref{fig:EDcompare}(b) shows the norm error \begin{equation}
\delta \psi = 1-\left|\langle\psi_{\mathrm{MPS}}|U_{\mathrm{ED}}|\psi_{\mathrm{MPS}}\rangle\right|
\end{equation}
comparing the MPS representation of the state with the exact one for varying bond dimension $\chi_{\mathrm{max}}$.  
Here $\psi_{\mathrm{MPS}}$ is the eigenstate created from contracting the truncated MPS along all the bond indices.
If $T_1$ is small, we obtain machine precision eigenstates already for $\chi_{\mathrm{max}}\approx 20$. 
For larger $T_1$, the required bond dimension saturates at the maximal value of $2^5=32$ (at which the full Hilbert space is spanned by the Schmidt states and DMRG is equivalent to ED).   
This behavior is hinting at the presence of an extended phase as the MPS description is no longer efficient. 
Lastly we choose a specific disorder realization and compare the energies $\theta_n$ of the DMRG-X algorithm initiated with different product states and ED results in Fig.~\ref{fig:EDcompare}(c) for $ T_1=0.2$.
As also observed in the original DMRG-X algorithm\cite{khemani2016obtaining}, certain energy levels are missing.
This happens if two or more eigenstates of $U_T$ have maximum weight on the same input basis state and can be avoided by requiring every new state to be orthogonal to the prior ones.
Further, for large $T_1$, missing quasi-energies and eigenstates may arise from a bias toward lower entropy eigenstates, as discussed in the concluding remarks below.
  
\section{ MBL beyond ED}\label{sec:big}
\begin{figure}
\begin{center}
\includegraphics[width=8cm]{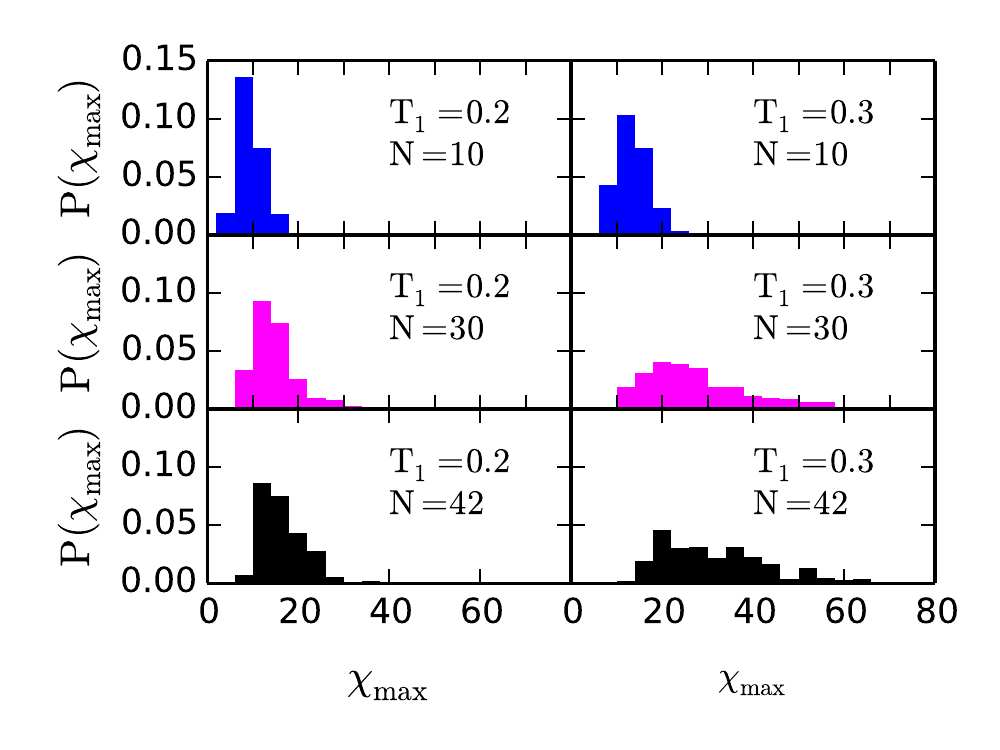}
\caption{Distribution of $\chi_{\mathrm{max}}$ necessary for $\mathrm{trunc}=10^{-12}$, taken from $\sim 300$ disorder realizations.  Since the distributions have been normalized to 1, they can be taken as probabilities of requiring a certain $\chi_{\mathrm{max}}$.  The left panel shows data for $T_1=0.2$ while the right panel shows data for $T_1=0.3$.\label{fig:histograms}}
\end{center}
\end{figure}

\begin{figure}
\begin{center}
\includegraphics[width=8cm]{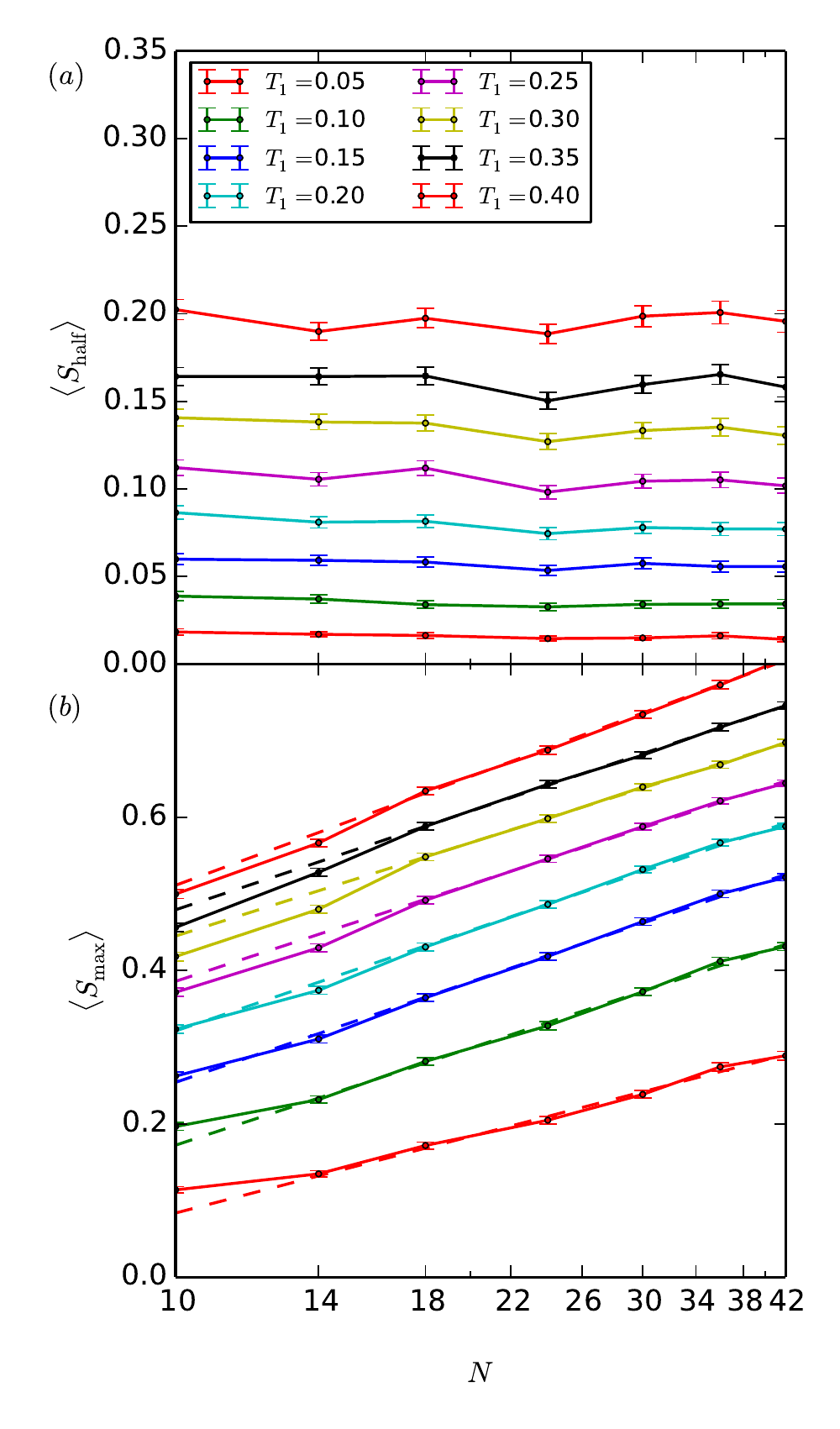}
\caption{(a) Disorder averaged half-chain entanglement entropy $\langle S_{\mathrm{half}}\rangle$. (b) Disorder averaged maximum entanglement over the bonds $\langle S_{\mathrm{max}}\rangle$ for $N=10-42$ and $T_1=0.05-0.40$, averaged over 400 disorder realizations. 
\label{fig:entropy}}
\end{center}
\end{figure}
We now go beyond the ED limit and consider systems up to $N=42$ sites.
In order to faithfully represent the Floquet operator, we choose, for each value of $T_1$, the $D_{\mathrm{max}}$ such that the truncation error is small compared to the Trotter error which is of order $10^{-8}$.
The more delicate part is the bond dimension $\chi_{\mathrm{max}}$ used in for the DMRG-X algorithm, since it varies with system size.
Moreover, as we iterate over different disorder realizations and eigenstates, we find that the required bond dimensions fluctuate strongly.
We thus choose the bond dimension dynamically by ensuring that the truncation error 
\[
\mathrm{trunc} = \displaystyle\sum_{\alpha=\chi_{\mathrm{max}}}^{d^{N/2}} \Lambda_{\alpha}^2
\]
at each DMRG step is smaller than $10^{-12}$. 
As independent measure we calculate  
\[
\delta E = 1-\left|\langle\psi_{\mathrm{MPS}}|U_{\mathrm{MPO}}|\psi_{\mathrm{MPS}}\rangle\right|
\]
to make sure that the eigenstates are well converged. 
We only show data with relatively small $T_1$ for which all simulations for different disorder realizations converged with a moderate bond dimension of $\chi_{\mathrm{max}} \le 52$. 
%
The histograms of $\chi_{\mathrm{max}}$'s for different system sizes and $T_1$ are shown in Fig.~\ref{fig:histograms}.
For $T_1=0.2$ (deep in the MBL phase), we find that $\chi_{\mathrm{max}}$ is basically independent of system size. 
However, already for $T_1=0.3$ we observe a noticeable a tail toward larger $\chi_{\mathrm{max}}$, which is accounted to the appearance of Griffith regions and resonances.
When going to larger $T_1$, which lie in the extended phase, the required $\chi_{\mathrm{max}}$ diverges quickly and cannot be simulated anymore (not shown). 

Figure~\ref{fig:entropy}(a) shows the half chain entanglement entropy for various system sizes, displaying a clear area law for small $T_1\le 0.4$.
The stability of the area law even for systems up to $N=42$ is a strong indication for the presence of an MBL phase. 
The disorder averaged {\it maximum} entanglement over the bonds $\langle S_{\mathrm{max}}\rangle$ however shows a logarithmic increase with system size.  
This increase is characteristic of the proliferation of Griffiths regions with anomalously low disorder.  The probability of a Griffiths region of size $L$ in a size $N$ chain scales as $Ne^{-L/L_0}$.  If we set this probability to a constant, then $L\propto\frac{1}{L_0}\log{N}$.  Since the entanglement entropy in these regions is proportional to their volume, the maximum entropy should indeed scale as $\log{N}$. This has also been discussed in the context of undriven MBL systems.\cite{khemani2016obtaining,Devakul16} We note that the Griffiths effects considered
here are in an MBL region. Presumably they connect across the localization transition to the Griffiths region discussed in 
Refs.~\onlinecite{agarwal2015anomalous,BarLev2015}.
%

\section{Conclusions}\label{sec:disco}
We have generalized the recently introduced Density-Matrix Renormalization Group for eXcited states (DMRG-X)\cite{khemani2016obtaining} algorithm to efficiently obtain Floquet eigenstates of one-dimensional, periodically driven many-body localized systems. 
To this end we introduced an algorithm to compress the Floquet operator in a matrix-product operator form of small bond dimension.
We tested the algorithm by considering a disordered spin-1/2 binary drive model. 
We first benchmarked the algorithm and confirmed that the algorithm reproduces the exact diagonalization results up to machine precision for small bond dimensions.  
We then considered system sizes that lie beyond exact diagonalization and focused on the regime in which the delocalizing time is short.
Here we find excellent convergence and observe an area law in the disorder averaged half chain entanglement entropy.
In addition, we find a logarithmic growth of the disorder averaged maximum entanglement entropy, which is due to Griffith regions.
For stronger driving (i.e. longer $T_1$), the system becomes more delocalized and with the spreading of resonant Griffiths regions, the assumption of local integrals of motion and small spin configuration change no longer holds.  
Here, DMRG-X  demonstrates a bias toward lower entropy eigenstates in the eigenstate selection step after energy targeting. 
This is due to lower entropy eigenstates typically having larger overlap with initial states that are zero entropy product states, and can be mollified by starting with random states beyond mere product states.
%
%
%
The point of onset of length-dependence, and thus delocalization and Griffiths regions, is not affected by eigenstate biasing because where Griffiths regions do not contribute significantly to entropy growth, there would be no length-dependence of entropy or other thermalization measures.  
Therefore, deep in the MBL phase, the system can be well described by the local integrals of motion picture and the DMRG-X method is not biased towards low entanglement states.
Going forward we expect this algorithm to be very useful in exploring the properties and stability of quantum orders in Floquet eigenstates. 

{\bf Acknowledgments.}
We thank I. Cirac for an especially enlightening discussion at the start of this project and V. Khemani for collaboration on much related work as well as discussions specific to this project.
SLS acknowledges support from the NSF-DMR via Grant No. 1311781, the Alexander von Humboldt Foundation via
a Humboldt Award. 
FP acknowledges support from the DFG (Deutsche Forschungsgemeinschaft) Research Unit FOR 1807 through grants no. PO 1370/2-1.

\bibliography{mbl}
\end{document}